\documentclass{IEEEconf}
\usepackage{cite}
\usepackage{graphicx}
\usepackage{listings}
\usepackage{caption}

\begin{document}

\title{Instantiating Standards: Enabling Standard-Driven Text TTP Extraction with Evolvable Memory}
\author{
Cheng Meng$^{1,2}$,ZhengWei Jiang$^{1,2}$,QiuYun Wang$^{1,2}$,XinYi Li$^{2}$,ChunYan Ma$^{1,2}$,FangMing Dong$^{1,2}$\\
FangLi Ren$^{1}$\thanks{FangLi Ren is the corresponding author}, BaoXu Liu$^{1,2}$\\[1ex]
$^{1}$Institute of Information Engineering, Chinese Academy of Sciences, Beijing, China\\
$^{2}$School of Cyber Security, University of Chinese Academy of Sciences, Beijing, China
}
\captionsetup{font=small}
\maketitle

\begin{abstract}
    Extracting MITRE ATT\&CK Tactics, Techniques, and Procedures (TTPs) from natural language threat reports is crucial yet challenging. Existing methods primarily focus on performance metrics using data-driven approaches, often neglecting mechanisms to ensure faithful adherence to the official standard. This deficiency compromises reliability and consistency of TTP assignments, creating intelligence silos and contradictory threat assessments across organizations. To address this, we introduce a novel framework that converts abstract standard definitions into actionable, contextualized knowledge. Our method utilizes Large Language Model (LLM) to generate, update, and apply this knowledge. This framework populates an evolvable memory with dual-layer situational knowledge instances derived from labeled examples and official definitions. The first layer identifies situational contexts (e.g., "Communication with C2 using encoded subdomains"), while the second layer captures distinctive features that differentiate similar techniques (e.g., distinguishing T1132 "Data Encoding" from T1071 "Application Layer Protocol" based on whether the focus is on encoding methods or protocol usage). This structured approach provides a transparent basis for explainable TTP assignments and enhanced human oversight, while also helping to standardize other TTP extraction systems. Experiments on a dataset show our framework (using Qwen2.5-32B) boosts Technique F1 scores by 18\% over GPT-4o. Qualitative analysis confirms superior standardization, enhanced transparency, and improved explainability in real-world threat intelligence scenarios. To the best of our knowledge, this is the first work that uses the LLM to generate, update, and apply the a new knowledge for TTP extraction.
\end{abstract}

\section{Introduction}

Cyber Threat Intelligence (CTI) involves the systematic collection, processing, and analysis of information pertaining to cyber threats, enabling organizations to bolster their defensive capabilities. \cite{sun2023cyber}Among various CTI sources, textual documents such as technical reports and online security articles are particularly rich in detailed threat information. \cite{rahman2023attackers}A critical objective in leveraging these textual sources is the accurate identification and comprehension of adversary operational methods.This granular understanding of operational methodologies allows organizations to transition from reactive awareness to proactive defense. Specifically, it facilitates the tailoring of security controls, prediction of potential attack vectors, recognition of malicious activity patterns, and ultimately, the optimized allocation of resources against observed adversary behaviors.

The MITRE ATT\&CK® framework provides a globally recognized standard for this purpose, cataloging adversary tactics, techniques, and procedures (TTPs) based on real-world observations. However, a significant challenge lies in mapping the unstructured, narrative descriptions within CTI sources to the structured entries of the ATT\&CK framework. Currently, this mapping relies heavily on manual analysis by security experts—a process that is inherently resource-intensive, time-consuming, and susceptible to inconsistencies. This reliance on manual TTP extraction consequently forms a bottleneck, hindering the timely operationalization of intelligence derived from textual CTI reports. Automating TTP extraction from CTI text using Natural Language Processing (NLP) techniques presents a viable solution. 

While existing TTP extraction systems using powerful NLP models often exhibit strong performance, they focus on fitting the dataset instead of following the official standard directly. This is not a problem if the dataset is consistent with the official standard. But, the continuous evolution of the ATT\&CK framework and the inherent preferences and biases of human experts, lead to inconsistencies in labeled data. Consequently, this severely undermines their reliability and practical applicability, leading to intelligence silos and contradictory threat assessments.

The challenge of rectifying this misalignment with the official standard, stemming from systems fitting inconsistent datasets, is significantly exacerbated by the black box nature of many methods. When discrepancies with the standard do emerge, this lack of transparency prevents clear diagnosis, hinders reconciliation efforts between different systems, and impedes trustworthy validation. In other words, while we can score performance on test data, the opacity of these methods makes fail to align with the standard's principles, hindering true trustworthiness and debuggability.

The challenges of misalignment with the official standard and the critical lack of transparency, underscore an urgent need for TTP extraction methods that are not only standard-adherent but also inherently transparent and explainable. While invaluable for human analysts, the current design of the official MITRE ATT\&CK framework presents specific obstacles to creating a truly faithful and interpretable automated system. These obstacles include:
\begin{itemize}
\item Firstly, its extensive scale, encompassing hundreds of techniques and sub-techniques, complicates the efficient and precise localization of specific candidate TTPs from textual data for automated systems.
\item Secondly, the framework lacks explicit delineation of contrastive relationships or potential overlaps between similar techniques, which is critical for automated disambiguation and accurate fine-grained classification.
\item Thirdly, the rich detail and narrative scope of TTP definitions, while excellent for comprehensive human understanding and threat analysis, render them too unwieldy and verbose for direct and efficient use in rapid, accurate machine-led classification.
\end{itemize}

Given these shortcomings of the official standard, this paper proposes a novel situational knowledge representation (SKR) as a foundational component for improved TTP extraction. This SKR acts as the medium for the actionable instantiation of official standards, thereby laying the groundwork for standard-driven TTP extraction.This SKR is founded upon the principle of hierarchical abstraction, separating knowledge into dual layers: a description of shared foundational elements across specific techniques, which aids in retrieval(e.g. Credential dumping or extraction from system), and a situational specific technique description building on the shared description to guide the TTP extraction process(e.g. T1003: Directly dumps credentials from memory or system, T1555: Extracts credentials from password storage locations). This structured approach not only bridges the crucial gap between semantic similarity and contextual relevance for more accurate and consistent TTP assignments but also intrinsically provides a transparent and auditable basis for verifiable and explainable TTP extraction outcomes

To manage and utilize this novel SKR, we employ an LLM-driven framework centered around an Evolvable Memory System. Within the system, individual memory entries consist of these SKR instances, which embody the actionable instantiations of the official standard. The system automates key operations on these memory entries, handles their initial generation and subsequent updating, highlighting the memory's evolvable nature. The automated generation process uses the LLM to synthesize these entries by combining official TTP definitions and labeled contextually similar sentences. This process ensures the resulting knowledge is both standard-aligned and richly contextualized. We separate the usage of the system into two steps: 1). Retrieve the shared description for classification, 2). Retrieve the situational specific description with classification result for check. step2 can also collaborate with other TTP extraction systems, helping to improve the consistency of other systems by using our framework to reclassify the result.

In total, our contributions are as follows:
\begin{itemize}
    \item \textbf{Novel Knowledge Representation for Standard Instantiation}: 
    We design and introduce a novel, dual-layer Knowledge Representation KR specifically tailored to instantiate the MITRE ATT\&CK standard for automated TTP classification tasks. This KR employs hierarchical abstraction by separating shared from situational-specific descriptions to overcome limitations of the raw standard, provide actionable guidance, and serve as the foundation for transparent, explainable TTP assignments. In technique views, the structure alleviate the contradiction between the semantic similarity and contextual relevance.
    \item \textbf{LLM-Driven Framework with Evolvable Memory System}: 
    We develop an LLM-driven framework featuring an Evolvable Memory System for the automated lifecycle management of the SKR, including its generation, refinement, and application. This framework synthesizes knowledge from official definitions and labeled similar sentences. The two-step usage of the framework achieve the accurate and consistent TTP extraction while be able to help other TTP extraction systems to improve their consistency.
    \item \textbf{Demonstrated Superiority in Standard-Driven TTP Extraction}: 
    We present extensive experiments validating our framework's effectiveness. Quantitative results show state-of-the-art TTP extraction performance, significantly surpassing strong baselines including retrieval-based methods and large commercial models like GPT-4o; for instance, our framework using Qwen2.5-32B achieved an 11\% higher F1 score than GPT-4o. Crucially, qualitative analysis and case studies confirm the framework's primary advantage: substantially improved consistency, transparency, and explainability in TTP assignments compared to conventional approaches, validating its trustworthiness for practical CTI analysis.
\end{itemize}

\section{Related Work}
The exploration of generative LLMs for identifying MITRE ATT\&CK TTPs from CTI reports has manifested in various research efforts, ranging from direct prompting of general-purpose models to the development of fine-tuned systems and Retrieval-Augmented Generation (RAG) systems.

Fayyazi and Yang (2023)\cite{fayyazi2023uses} conducted a study evaluating the use of GPT-3.5 (specifically gpt-3.5-turbo) and Google's Bard for interpreting ambiguous cyberattack descriptions and mapping them to corresponding MITRE ATT\&CK tactics.

Mezzi et al. (2025)\cite{mezzi2025largelanguagemodelsunreliable} conducted an evaluation that included fine-tuning state-of-the-art LLMs—specifically gpt4o, gemini-1.5-pro-latest, and mistral-large-2—on CTI tasks, one of which was attack vector extraction (analogous to TTP extraction). Using a dataset of 350 real-size Advanced Persistent Threat (APT) reports from MITRE, structured in STIX format, their findings on the efficacy of fine-tuning were mixed. While Gemini demonstrated the best F1-score (0.87) for attack vector extraction after fine-tuning, gpt4o and Mistral did not exhibit significant improvements (both achieving an F1-score of 0.67). Furthermore, the study noted that fine-tuning could sometimes worsen the confidence calibration of the models. These empirical results provide crucial data on the actual effectiveness (or lack thereof) of fine-tuning prominent generative LLMs for TTP extraction from realistic CTI reports, suggesting that fine-tuning is not a universally guaranteed solution for these complex models and tasks.  

\section{Method}
The core of our framework is to instantiate the MITRE ATT\&CK standard to operational knowledge for classification task implemented through three key components: Situational Knowledge Representation, Life-long Memory Management and Standard-Driven TTP Extraction. 

\subsection{Overview of the Framework}
Our framework is designed to transform the abstract definitions of the MITRE ATT\&CK standard into actionable, contextualized knowledge, enabling more accurate, consistent, and transparent TTP extraction. At its heart, the framework leverages Large Language Model (LLM) to manage an Evolvable Memory System populated with instances of our novel Situational Knowledge Representation (SKR).

\begin{figure}[h]
    \centering
    \includegraphics[width=0.9\linewidth]{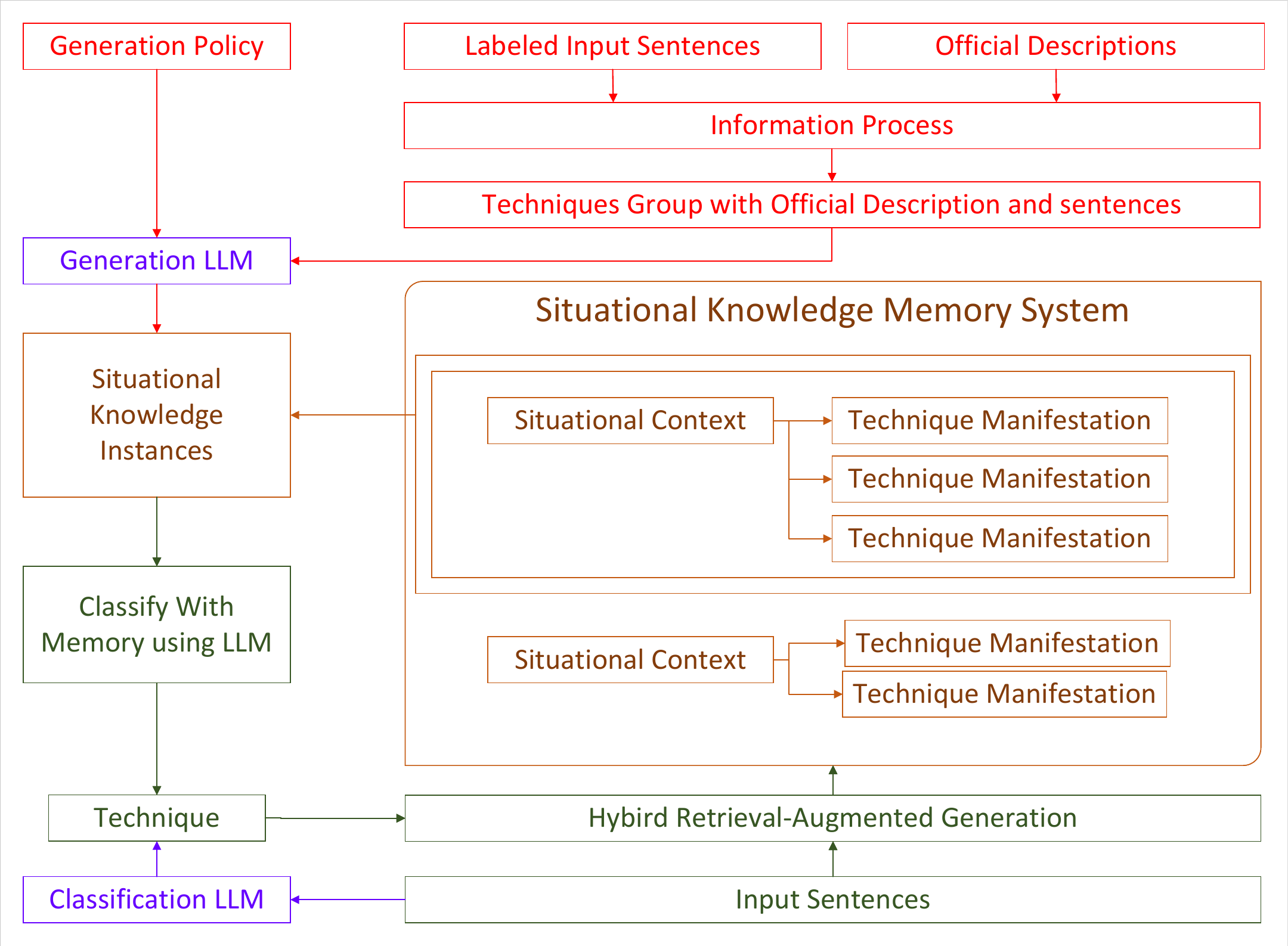}
    \caption{Overview of the main components of the framework, including Situational Knowledge with memory system, knowledge generation and two-step knowledge-driven TTP extraction.}
    \label{fig:method_overview}
\end{figure}

The process begins with the \textbf{Situational Knowledge Representation (SKR)}. This dual-layer representation is specifically designed to make the official standard more amenable to automated processing. Its hierarchical structure addresses the verbosity and lack of explicit contrastive relationships in the original ATT\&CK definitions.

Next, the \textbf{Life-long Memory Management} component governs the lifecycle of these SKR instances, supports the initial generation, continuous update and forget mechanism. LLM is employed to automatically generate initial SKR entries by synthesizing information from labeled textual examples from CTI reports and relevant official ATT\&CK definitions. This ensures that the knowledge is both standard-aligned and empirically grounded. 

Finally, the \textbf{Standard-Driven TTP Extraction} component applies the knowledge stored in the evolvable memory to identify TTPs in new, unseen CTI texts. This is achieved through a two-step process:
\begin{enumerate}
    \item Initial Retrieval and Classification: Given an input text, the system first retrieves relevant layer 1 from the memory to narrow down potential TTP candidates, then use layer 2 to guide the classification.
    \item Refinement and Verification: Then, the system retrieves the layer 2 with the classification result and input text to verify the classification result, given result in initial classification may ignore some specific details. This step can also be used to re-evaluate and standardize the outputs of other TTP extraction systems.
\end{enumerate}
Through these interconnected components, our framework aims to provide a transparent and auditable process for TTP extraction, alleviate the challenges of standard adherence and explainability that plague existing methods. The subsequent sections will detail each of these components.

\subsection{Situational Knowledge Representation (SKR)}

The core design philosophy of the SKR is a hierarchical, dual-layer architecture for organizing and articulating knowledge pertinent to ATT\&CK techniques. This structure emulates the cognitive process of human experts, who typically first identify a general attack scenario or intent and then differentiate similar techniques based on specific details. Through this approach, the SKR not only more accurately instantiates the official standard but also effectively addresses the challenges encountered when directly applying the standard's raw form to automated systems.
\begin{figure}[h]
    \centering
    \includegraphics[width=0.8\linewidth]{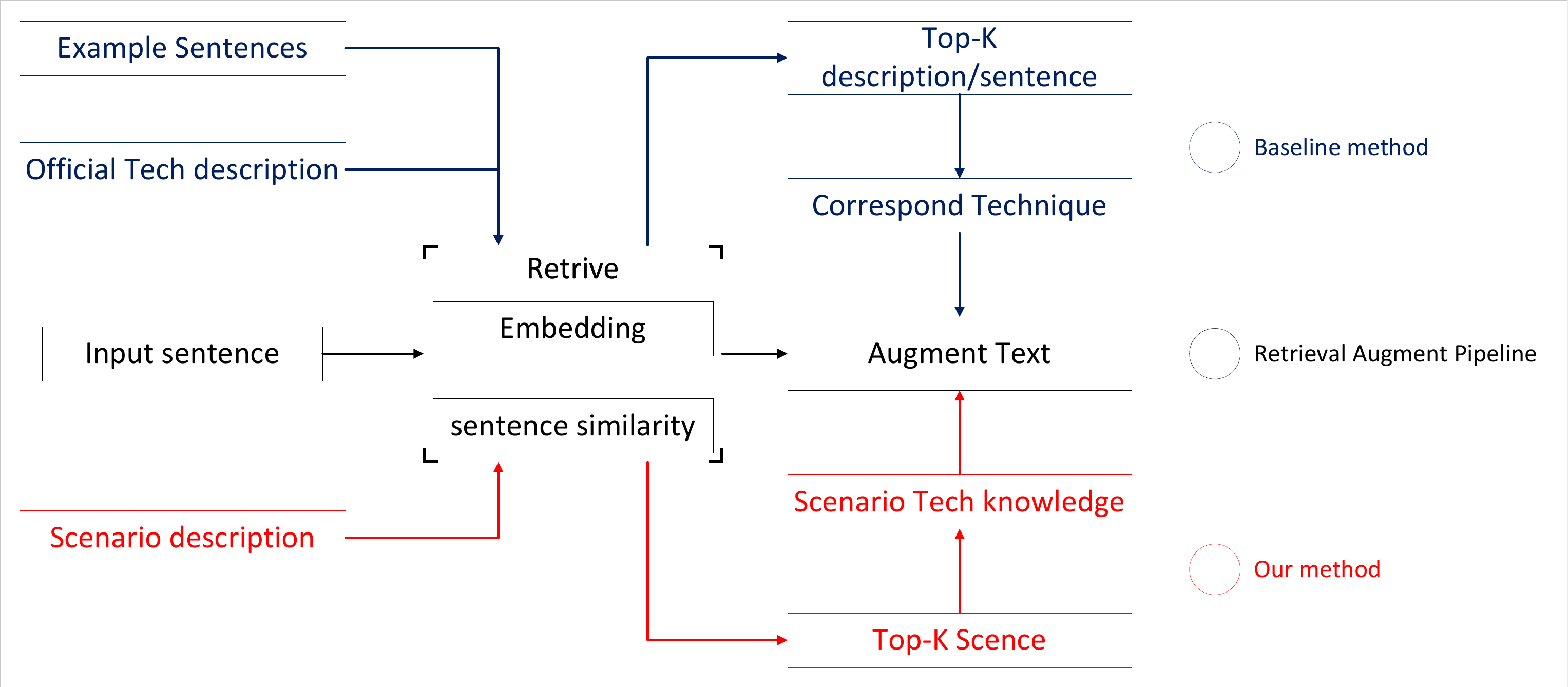}
    \caption{Key difference between the proposed Situational Knowledge Representation (SKR) and the other knowledge representation methods. Our SKR is a dual-layer knowledge representation, separate the retrieval text and classification guidance, thus alleviate the contradiction between the semantic similarity and contextual relevance.}
    \label{fig:classify_comparison}
\end{figure}
Each SKR instance is composed of two fundamental layers:
\begin{enumerate}
\item \textbf{Layer 1: Situational Context}
This layer aims to describe a relatively generalized, shared foundational behavior or attack scenario. It represents a common basis for a group of ATT\&CK techniques that are semantically related or functionally similar within the attack lifecycle. For example, "Credential dumping or extraction from system" or "Communication with C2 using encoded subdomains" are typical Situational Contexts.
\begin{itemize}
\item \textbf{Purpose}: To identify and categorize a specific class of adversary activity or intent, serving as an entry point and index for subsequent precise technique classification. It assists the system in rapidly localizing a relevant subset from the broad spectrum of ATT\&CK techniques.
\item \textbf{Characteristics}: The Situational Context is an abstract description of commonalities across multiple specific techniques, may corresponding to a general strategy an attacker might employ or a common situation they might encounter at a certain stage.
\item \textbf{Function}: In the TTP extraction process, this layer is primarily utilized for initial filtering and narrowing down the scope of candidate techniques, thereby enhancing the efficiency of subsequent fine-grained analysis.
\end{itemize}

\item \textbf{Layer 2: Specific Technique Manifestation}
Building upon a given Situational Context, this layer provides further granularity by capturing and describing the unique characteristics and key differentiators of individual ATT\&CK techniques within that context. It focuses on the specific manifestations or implementation details that clearly distinguish one technique from other similar ones. For instance, within the "Communication with C2 using encoded subdomains" Situational Context, the specific descriptions for different techniques are as follows:
\begin{itemize}
    \item The manifestation for \texttt{T1132} (\textit{Data Encoding}) is: "Uses base32 encoding for subdomains to obfuscate C2 communication," emphasizing the encoding method.
    \item The manifestation for \texttt{T1071} (\textit{Application Layer Protocol}) is: "Employs DNS as an application layer protocol for C2 communication," highlighting the protocol's usage.
\end{itemize}
The idea of this layer are as follows:
\begin{itemize}
    \item \textbf{Purpose}: To furnish a concise, actionable definition for each specific ATT\&CK technique, highlighting its core behavior and its distinctions from other techniques within a particular context.
    \item \textbf{Characteristics}: These descriptions are targeted and discriminative, designed to convert abstract technical definitions into concrete, observable or inferable indicators within actual textual data.
    \item \textbf{Function}: This layer provides a direct basis for the final TTP classification, ensuring accuracy and consistency. It also underpins the explainability of the assignments, as the decision-making process can be traced back to these specific, differentiating feature descriptions.
\end{itemize}
\end{enumerate}

This dual-layer structure of the SKR directly addresses the challenges associated with the automated application of the official ATT\&CK standard:
\begin{itemize}
\item \textbf{Addressing Extensive Scale}: The first layer, Situational Context, simplifies navigation and initial screening within the extensive set of techniques by grouping them.
\item \textbf{Explicitly Differentiating Similar Techniques}: The second layer, Specific Technique Manifestation, clarifies the boundaries and potential overlaps among similar techniques by providing contrasting descriptions within a shared context, thereby compensating for this deficiency in the official standard.
\item \textbf{Overcoming Verbose Definitions}: The SKR transforms the rich, and at times verbose, narrative definitions of the official standard into concise, actionable instantiations of knowledge optimized for classification.
\end{itemize}

The concrete representation of an SKR instance, as exemplified below, typically takes the form of a structured object. This object comprises a "state" field (corresponding to the first-layer Situational Context) and an "action" field (an object mapping TTP IDs to the second-layer Specific Technique Manifestation descriptions). For example:

\begin{lstlisting}[breaklines=true, basicstyle=\footnotesize\ttfamily]
{
  "state": "Communication with C2 using encoded subdomains",
  "action": {
    "T1132": "Uses base32 encoding for subdomains to obfuscate C2 communication",
    "T1071": "Employs DNS as an application layer protocol for C2 communication",
    "T1001": "Involves data obfuscation techniques like AES ciphertext within subdomains",
    "T1008": "Indicates fallback to alternative protocols like HTTP if primary DNS fails"
  }
}
\end{lstlisting}

Through this meticulously designed SKR structure, our framework can more effectively apply the ATT\&CK standard to automated TTP extraction tasks, while simultaneously enhancing the transparency of the process and the explainability of the results, thereby supporting the development of genuinely standard-driven threat intelligence analysis capabilities.

\subsection{Life-long Memory Management}
The Life-long Memory Management component is central to the framework's adaptability and continuous improvement. It orchestrates the entire lifecycle of Situational Knowledge Representation (SKR) instances within the Evolvable Memory System. This lifecycle encompasses three primary operations: the generation of memory entries, their subsequent refinement based on new information and performance feedback, and a forgetting mechanism to prune outdated or ineffective knowledge. The key is the generation of memory entries.

\subsubsection{Memory Generation}
The generation of memory entries, i.e., SKR instances, is a foundational process that populates the Evolvable Memory System with actionable knowledge derived from the MITRE ATT\&CK standard and empirical data. This process is primarily driven by a Large Language Model (LLM), tasked with synthesizing robust and contextualized SKR instances.

The population of the memory involves the automated creation of SKR instances from a combination of diverse information sources. The LLM serves as the core engine for this synthesis. The key inputs to the LLM for generating a single SKR instance include:
\begin{itemize}
\item Target Sentence from CTI Report: A specific sentence or snippet from a cyber threat intelligence report that describes an adversary behavior and is labeled with an ATT\&CK Technique (ground truth). This provides concrete, real-world examples of how techniques manifest.
\item Official ATT\&CK Definitions: The official names and descriptions of the ATT\&CK techniques relevant to the target sentence and its labels. This ensures that the generated knowledge remains anchored to the standard.
\item Contextually Similar Sentences: To enrich the context and help the LLM generalize, the system performs a semantic search against a knowledge base of existing CTI sentences. This search retrieves a set of sentences (e.g., top 5) that are semantically similar to the target sentence and are also tagged with relevant TTPs. These similar cases offer varied phrasings and contexts for related behaviors. 
\end{itemize}
The difference between the initialization and the update is that the initialization use all Target Sentences as the Contextually Similar Sentences, and the update use the old Target Sentences recorded in the Memory System as the Contextually Similar Sentences.

The generation process, as implemented in our framework , structures these inputs into a carefully designed prompt for the LLM. The prompt guides the LLM to perform the following:
\begin{itemize}
\item Identify a "State" (Layer 1 of SKR): Based on the target sentence, its associated TTPs, the official descriptions, and the similar sentences, the LLM formulates the situational context. This "state" description is designed to be technique-agnostic yet specific enough to capture a class of adversary activity (e.g., "Communication with C2 using encoded subdomains"). It is designed to be broad enough to potentially encompass multiple related techniques.
\item Define "Actions" (Layer 2 of SKR): For each relevant ATT\&CK technique ID identified in the input (either from the target sentence's labels or the similar sentences), the LLM generates a concise, discriminative description. This "action" explains the key distinguishing features of that specific technique within the context of the generated "state." It clarifies why the observed behavior points to this particular technique and how it differs from other similar techniques that might fall under the same "state." For example, for T1132, the action might be "Uses base32 encoding for subdomains to obfuscate C2 communication," while for T1071, it might be "Employs DNS as an application layer protocol for C2 communication." Example sentences from the similar data can also be incorporated into the action description to provide concrete illustrations.
\end{itemize}

\subsubsection{Memory Optimization}
Memory Optimization primarily expands the existing knowledge base by adding new "actions" (Layer 2 of SKR) to Situational Knowledge Representation (SKR) instances, rather than modifying existing components. Both the established "state" (Layer 1 of SKR) and its pre-existing "actions" are considered robust and are preserved, based on trust in the knowledge generation strategy. When new textual evidence indicates a technique manifestation not yet covered under a particular "state, the optimization process focuses exclusively on creating and incorporating new technique-specific "actions" linked to the relevant, established "state." 

A Large Language Model (LLM) facilitates this by generating these new "action" descriptions tailored to the original memory and the new textual evidence. The LLM also assists in resolving potential conflicts if multiple distinct new actions are proposed for the same state-technique pairing. This approach incrementally extends the SKR's coverage for a wider array of techniques within recognized behavioral contexts, without altering the validated core knowledge of existing states and actions.
\subsubsection{Memory Forget}
Memory Forgetting is crucial for maintaining the long-term relevance and efficiency of the Evolvable Memory System. This mechanism systematically prunes SKR instances that consistently underperform or lead to incorrect Technique assignments. The process involves tracking the classification accuracy associated with each memory entry, specifically how its "state" and consequent "actions" contribute to outcomes. A scoring mechanism evaluates this performance against predefined criteria. Entries that fall below a utility threshold, indicating low effectiveness or a detrimental impact on overall accuracy, are removed. This ensures the memory remains compact, current, and populated with high-utility knowledge, preventing degradation from obsolete or misleading entries.

\subsection{Standard-Driven TTP Extraction}
The Standard-Driven TTP Extraction component operationalizes the knowledge stored within the Evolvable Memory System to analyze new Cyber Threat Intelligence (CTI) texts. This process systematically assigns MITRE ATT\&CK Tactics, Techniques through a two-stage methodology. The first stage performs an initial retrieval and classification. The second stage then refines and verifies a given TTP classification, which can be the output of our first stage or the result from an external TTP extraction system, highlighting our framework's capability to assist and standardize other systems.

The framework's two-stage approach provides robust TTP extraction. The first stage delivers an initial, contextually-aware classification by leveraging both layers of the SKR. The second stage offers a focused refinement capability, particularly valuable for disambiguating complex cases or standardizing inputs from various sources, by concentrating on the contrastive details within the Specific Technique Manifestations. The entire framework design supports scalability through concurrent processing capabilities.

\subsubsection{Stage 1: Initial Retrieval and Classification}

This initial stage is designed to identify a relevant subset of TTPs for a given input CTI text snippet and to produce an initial TTP classification. The process unfolds as follows:

\textbf{Memory Retrieval based on Situational Context (Layer 1)}: Upon receiving an input CTI text, the system first queries the Evolvable Memory System. A semantic retrieval mechanism is employed to fetch SKR instances whose Situational Context (Layer 1) aligns with the adversarial behavior described in the input text. 

\textbf{LLM-based Initial Classification guided by Retrieved Knowledge}: We retrieve k SKR instances from the memory by sentence similarity. A prompt is then dynamically constructed for the Large Language Model (LLM). This prompt integrates:
\begin{itemize}
\item The input CTI text.
\item The Situational Context (Layer 1) from each of the k selected SKR instances, providing the LLM with the overarching operational context.
\item The Specific Technique Manifestation (Layer 2) associated with each Situational Context. This supplies the LLM with concise, discriminative features and illustrative examples that differentiate the TTPs applicable within those contexts.
\item A list of candidate TTP IDs, derived from the TTPs detailed within the Specific Technique Manifestation entries of the selected k SKR instances.
\end{itemize}
The LLM processes this structured prompt to generate an initial TTP classification for the input text. The system supports efficient batch processing of multiple CTI text snippets through parallelization to handle larger datasets.

\subsubsection{Stage 2: Refinement and Verification of TTP Classification}

The second stage is designed to refine and verify a given TTP classification, enhancing its accuracy and consistency by explicitly considering fine-grained distinctions between similar techniques. A key feature of this stage is its ability to process TTP classifications originating either from Stage 1 of our own framework or from external TTP extraction systems. This flexibility allows our framework to serve as a verification and standardization layer, thereby improving the outputs of other systems.

\textbf{Targeted Memory Retrieval for Disambiguation}: This step takes the input CTI text and a TTP classification. Based on this input TTP classification, the system performs a targeted retrieval or filtering of SKR instances. Specially, system retrieve the Specific Technique Manifestation rather than Situational Context used in Stage 1. The goal is to gather SKR instances that are particularly relevant for disambiguating the input TTP classification from other closely related or potentially confusable techniques. 
\textbf{Contrastive Prompting and LLM-based Re-evaluation/Re-classification}: A specialized prompt is constructed for the LLM for a nuanced re-evaluation. Distinct from the Stage 1 prompt, this version integrates the prior classification result and is revised to better highlight distinguishing characteristics relevant to the re-classification.

The LLM is then tasked to re-classify or verify the text based on this detailed and contrastive information. This refined classification benefits from the explicit articulation of differentiating knowledge from Layer 2 of the SKR, enabling the model to resolve ambiguities and either confirm or correct the input TTP assignment with greater confidence.

\section{Evaluation}
We evaluate the framework on the dataset provided by{\cite{nguyen-etal-2024-noise}}, and the result is shown in the following tables. SKR1 means our method with only classification stage1, SKR2 means our method with classification stage1 and verification stage2. Base means use the text without any additional information. official means use the official MITRE ATT\&CK TTP definitions. Procedures means use the procedures of the TTP definitions, which seems like to be cheating, but we still include it for comparison. All the result use the same model without memtioned, which is Qwen2.5-32B. GPT-4o and Deepseek-v3 use no additional information.

\begin{table}[h]
    \centering
    \caption{Evaluation on Technique(resolving sub-technique to its parent technique) extraction systems on procedures dataset}
    \label{tab:result comparison with training}
    \begin{tabular}{lcccc}
        \hline
        System & Acc & Prec & Rec & F1 \\
        \hline
        Base & 0.30 & 0.12 & 0.12 & 0.10 \\
        official & 0.58 & 0.49 & 0.55 & 0.47 \\
        procedures & 0.62 & 0.40 & 0.37 & 0.36 \\
        SKR1 & 0.75 & 0.63 & 0.59 & 0.59 \\
        SKR2 & 0.78 & 0.67 & 0.64 & 0.63 \\
        GPT-4o & 0.66 & 0.51 & 0.46 & 0.45 \\
        Deepseek-v3 & 0.67 & 0.49 & 0.48 & 0.46 \\
        \hline
    \end{tabular}
\end{table}

\begin{table}[h]
    \centering
    \caption{Evaluation on Technique(resolving sub-technique to its parent technique) extraction systems on expert dataset}
    \label{tab:result comparison without training}
    \begin{tabular}{lcccc}
        \hline
        System & Acc & Prec & Rec & F1 \\
        \hline
        Base & 0.16 & 0.07 & 0.10 & 0.07 \\
        official & 0.37 & 0.32 & 0.35 & 0.30 \\
        procedures & 0.36 & 0.27 & 0.29 & 0.25 \\
        SKR1 & 0.39 & 0.31 & 0.36 & 0.31 \\
        SKR2 & 0.41 & 0.36 & 0.41 & 0.35 \\
        GPT-4o & 0.23 & 0.16 & 0.18 & 0.15 \\
        \hline
    \end{tabular}
\end{table}
Notice, the 4o result use part of the test set, we will finish soon.

\section{Limitations}
To make the framework more reliable, Here are some future works:
\begin{itemize}
    \item Knowledge representation is still a problem. What the situation context represent should be more specific.
    \item A embedding model specially trained or fine-tuned for graping the similarity between the attack behavior description is needed.
    \item A more powerful and low-cost general model is needed to support the framework.
    \item Data-driven: The framework is still data-driven, which means the performance of the framework is limited by the quality of the data.
\end{itemize}

\section{Conclusion}
Our work propose a novel framework for TTP extraction from CTI reports, which is standard-driven using the Situational Knowledge Representation (SKR) and the Life-long Memory Management. To the best of our knowledge, this is the first work that generate the new form information to support the TTP extraction. The framework is evaluated on the dataset provided by{\cite{nguyen-etal-2024-noise}}, and the result shows that the framework is effective and efficient.

\bibliographystyle{IEEEtran}
\bibliography{reference}

\end{document}